\documentclass[12pt]{article}
\usepackage{authblk}
\usepackage[bookmarksnumbered, colorlinks=false, plainpages, hidelinks]{hyperref}
\usepackage{amsmath, amsthm, amscd, amsfonts, amssymb, graphicx, color, booktabs, sectsty, geometry}
\sectionfont{\fontsize{14}{15}\selectfont}
\numberwithin{equation}{section}
\definecolor{email}{rgb}{0.00,0.00,0.84}
\begin{document}
\setcounter{page}{1}

\title{\large \bf 12th Workshop on the CKM Unitarity Triangle\\ Santiago de Compostela, 18-22 September 2023 \\ \vspace{0.3cm}
\LARGE Search for CP violation in baryons with the LHCb detector}


\author{Chiara Mancuso\textsuperscript{1,2,3} on behalf of the LHCb collaboration \\
        \textsuperscript{1}Université Paris-Saclay, CNRS/IN2P3, IJCLab, Orsay, France \\
        \textsuperscript{2}INFN Sezione di Milano, Milano, Italy \\
        \textsuperscript{3}Università degli Studi di Milano, Milano, Italy }

\maketitle

\begin{abstract}
The latest results from the LHCb collaboration for the search for \textit{CP} violation in $b$-baryon decays are reported here. The first article presented is the search conducted in the $\Lambda_b^0 \rightarrow p \pi^- \pi^+ \pi^- $ decay, and the subsequent observation of \textit{P} violation. The following paper describes the search and the first observation of the $\Lambda_b^0 \rightarrow D^0 p K^-$ decay, with $D^0 \rightarrow K \pi$, Cabibbo-favoured and doubly suppressed mode. This proceeding concludes with the first amplitude analysis performed in $\Xi_b^- \rightarrow p K^- K^-$, whose model allows for \textit{CP} violation.
\end{abstract} \maketitle

\section{Introduction}

\noindent Studying \textit{CP} violation is fundamental to strengthen the Standard Model (SM) but also to challenge its weaknesses. 
Within the framework of the SM, \textit{CP} violation is accommodated by the Cabibbo-Kobayashi-Maskawa (CKM) matrix \cite{PhysRevLett.10.531, 10.1143/PTP.49.652}, which describes in weak interactions, the mixing between different quark flavours and allows for a complex phase, thus providing a source for \textit{CP} violation. The CKM matrix offers a coherent explanation for the observed \textit{CP} violating effects, especially in the sector of mesons. \textit{CP} violation in the $K$ \cite{PhysRevLett13138}, $B$ \cite{PhysRevLett.86.2515, Abe_2001} and $D$ \cite{PhysRevLett.122.211803} meson systems, is well-established experimentally and is consistent with the predictions of the SM.
Despite the successes of the CKM matrix in explaining \textit{CP} violation in mesons, the baryon sector remains a largely uncharted territory. Indeed, the prediction of \textit{CP} violation in baryons has never been experimentally confirmed. In this context, the most recent efforts of the LHCb collaboration are reported in the following.

\section{Search for \textit{CP} violation and observation of \textit{P} violation in $\Lambda_b^0 \rightarrow p \pi^- \pi^+ \pi^-$}
The first study presented describes a detailed analysis of the decay $\Lambda_b^0 \rightarrow p \pi^- \pi^+ \pi^-$ \cite{PhysRevD.102.051101}, using the data obtained from proton-proton ($pp$) collisions, with an integrated luminosity of 6.6 fb$^{-1}$ collected over the period from 2011 to 2017 at energies of $\sqrt{s}$ = 7, 8, and 13 TeV. The measurement employs two distinct and independent techniques: the triple product asymmetries and the unbinned energy test method. 

The searches for \textit{CP} violation are performed by separating \textit{P}-odd and \textit{P}-even contributions. These investigations rely on the utilization of a rich control sample, the Cabibbo-favoured $\Lambda_b^0 \rightarrow \Lambda_c^+(\rightarrow p K^- \pi^+) \pi^-$ decay, expected to exhibit no \textit{CP} violation and is employed to evaluate potential experimental biases and systematic effects.

The scalar triple products are built in the $\Lambda_b^0$ rest frame and correspond to
$C_{\widehat{T}} \equiv \vec{p}_p \cdot\left(\vec{p}_{\pi_{\text {fast }}^{-}} \times \vec{p}_{\pi^{+}}\right)$ and $\bar{C}_{\widehat{T}} \equiv \vec{p}_{\bar{p}} \cdot\left(\vec{p}_{\pi_{\text {fast }}^{+}} \times \vec{p}_{\pi^{-}}\right)$; by definition these are odd under $\widehat{T}$ operator transformations, which reverses momentum and spin three-vectors. \textit{CP} and \textit{P} violating effects appear as differences between the triple product observables related by \textit{CP} and \textit{P} transformations. Indeed, the asymmetries are defined as
\begin{align*}
    A_{\widehat{T}} &= \frac{N\left(C_{\widehat{T}}>0\right)-N\left(C_{\widehat{T}}<0\right)}{N\left(C_{\widehat{T}}>0\right)+N\left(C_{\widehat{T}}<0\right)} & \bar{A}_{\widehat{T}} &= \frac{\bar{N}\left(-\bar{C}_{\widehat{T}}>0\right)-\bar{N}\left(-\bar{C}_{\widehat{T}}<0\right)}{\bar{N}\left(-\bar{C}_{\widehat{T}}>0\right)+\bar{N}\left(-\bar{C}_{\widehat{T}}<0\right)} ,
\end{align*}
where $N$ and $\bar{N}$ are the number of events of $\Lambda_b^0$ and $\bar{\Lambda_b^0}$.
It follows that the \textit{CP} and \textit{P} violating asymmetries are
\begin{align*}
a_{C P}^{\widehat{T} \text {-odd }} & =\frac{1}{2}\left(A_{\widehat{T}}-\bar{A}_{\widehat{T}}\right) & 
a_P^{\widehat{T} \text {-odd }} &= \frac{1}{2}\left(A_{\widehat{T}}+\bar{A}_{\widehat{T}}\right).
\end{align*}

\begin{figure}
    \centering
    \includegraphics[scale = 0.45]{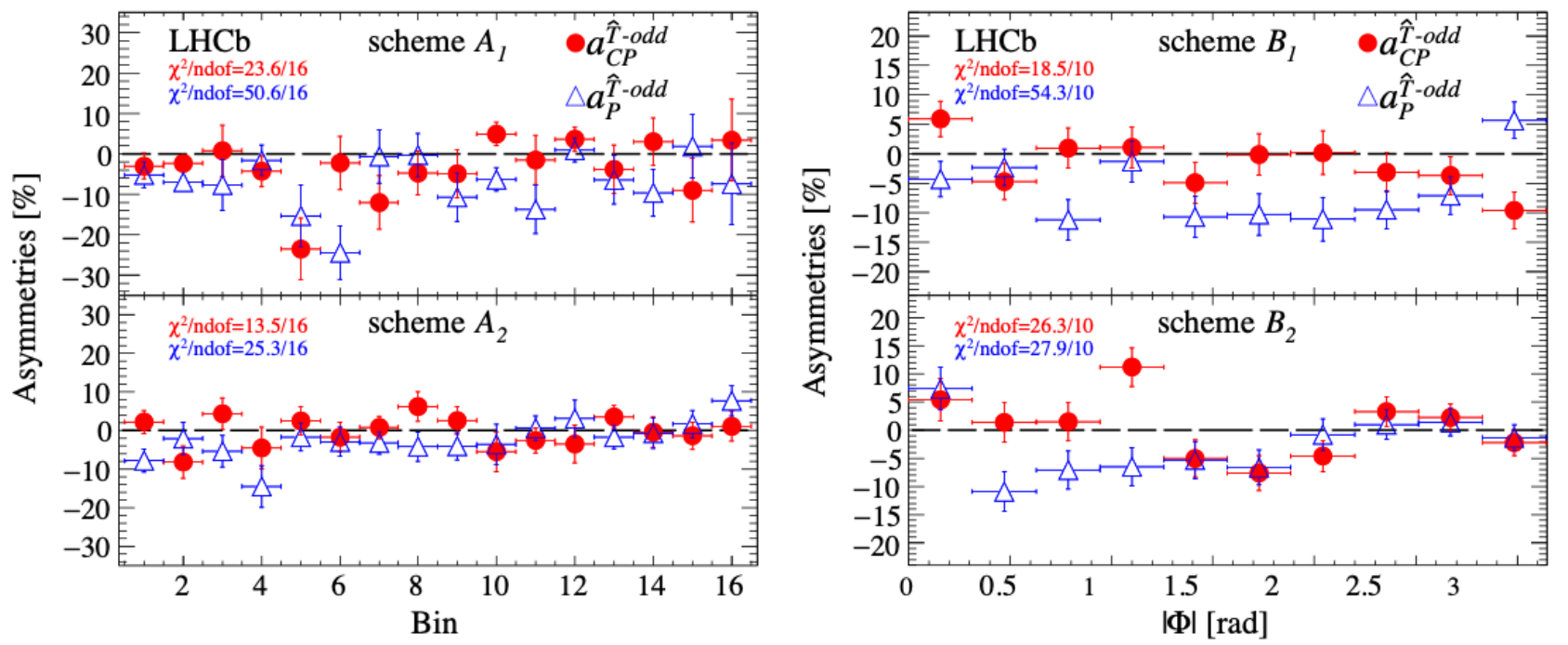}
    \caption{Measured asymmetries for the binning scheme (left) $A_1$ and $A_2$ and (right) $B_1$ and $B_2$. The error bars represent the sum in quadrature of the statistical and systematic uncertainties. The $\chi^2$ per ndof is calculated with respect to the null hypothesis and includes statistical and systematic uncertainties.}
    \label{fig:binnedTPA}
\end{figure}

These quantities have been measured in the full phase space but also in specific regions, as the decay investigated has a rich resonant structure whose main contributions are given by $N^{*+} \rightarrow \Delta^{++}(1234) \pi^{-}$, with $\Delta^{++}(1234) \rightarrow p \pi^{+}$ and $a_1^{-}(1260) \rightarrow \rho^0(770) \pi^{-}$, with $\rho^0(770) \rightarrow \pi^{-} \pi^{+}$. These asymmetries are designed to be substantially unaffected by asymmetries arising from particle-antiparticle production and those induced by the detector. To enhance the sensitivity towards localized \textit{CP} violation effects, the phase space is segmented into bins following two distinct binnings: the first one, labeled as A, divides the dataset into 16 subsamples to explore the distribution of the polar
and azimuthal angles of the proton in the $\Delta^{++}$ rest frame; the second one, labeled as B, divides the dataset into 10 subsamples, used to probe the asymmetries as a function of $|\Phi|$, the angle between the planes defined by the $p \pi^-_{fast}$ and $\pi^+ \pi^-_{slow}$ systems in the $\Lambda_b^0$ rest frame.
The results for the full phase space are the following
\begin{align*}
a_{C P}^{\widehat{T} \text {-odd }} & = (-0.7 \pm 0.7 \pm 0.2)\% & 
a_P^{\widehat{T} \text {-odd }} &= (-4.0 \pm 0.7 \pm 0.2) \%.
\end{align*}
The value for obtained for $a_{C P}^{\widehat{T} \text {-odd }}$ is in agreement with the \textit{CP} conserving hypothesis, whereas the $a_P^{\widehat{T} \text {-odd }}$ is clearly far from zero. Through the profile likelihood-ratio test, it is found that the value has a significance level of 5.5 $\sigma$, suggesting a violation of parity in this channel. Figure \ref{fig:binnedTPA} shows the results for the binned schemes $A$ and $B$ in two invariant mass regions: in $A_1$ and $B_1$, the schemes are applied in the region where the resonance $a_1$ is dominant, i.e. $m(p \pi^+ \pi^-_{slow}) > 2.8$ $\text{GeV}/c^2$, while $A_2$ and $B_2$ analyse the area in which the $N^{*+}$ resonance is dominant, i.e. $m(p \pi^+ \pi^-_{slow}) < 2.8$ $\text{GeV}/c^2$.

The energy test is a model-independent unbinned test sensitive to local differences between two samples, which would be the case for \textit{CP} violation. The test is performed through the calculation of a test statistic
\begin{equation}
    T = \frac{1}{2n(n-1)} \sum_{i \neq j}^{n} \psi_{ij} + \frac{1}{2\bar{n}(\bar{n}-1)} \sum_{i \neq j}^{\bar{n}} \psi_{ij} - \frac{1}{n\bar{n}} \sum_{i=1}^{n} \sum_{j=1}^{\bar{n}} \psi_{ij},
\end{equation}
where there are $n$ ($\bar{n}$) candidates in the first (second) sample. The first (second) term sums over pairs of candidates drawn from the first (second) sample and the final term sums over pairs with one candidate drawn from each sample. Each pair of candidates $ij$ is assigned a weight $\psi_{ij} = e^{-d_{ij}^2/2\delta^2} $, where $d_{ij}$ is their Euclidean distance in phase space, while the tunable parameter $\delta$ determines the distance scale probed using the energy test. The phase space is defined using the squared masses \( m^2(\pi^+), m^2(\pi^+\pi^-_{\text{slow}}), m^2(p \pi^+\pi^-_{\text{slow}}\pi^-_{\text{fast}}) \), \( m^2(\pi^-\pi^+_{\text{slow}}) \) and \( m^2(p \pi^-_{\text{slow}}) \). The value of \( T \) is large when there are significant localized differences between samples and has an expectation of zero when there are no differences. The distribution of \( T \) under the hypothesis of no sample differences, and the assignment of $p$-values, are determined using a permutation method. Results for different configurations of the energy test are summarized in Table \ref{tab:pvalue}. All \textit{CP} violation searches using the energy test result in $p$-values with a significance of 3$\sigma$ or smaller. The \textit{P} violation test shows a significance of 5.3 $\sigma$ at the two smaller scales probed.

\begin{table}[h!]
\centering
\small
\begin{tabular}{lccc}
\hline
\hline
Distance scale $\delta$ & 1.6 GeV$^2/c^4$ & 2.7 GeV$^2/c^4$ & 13 GeV$^2/c^4$ \\
\hline
$p$-value (CP conservation, P even) & $3.1 \times 10^{-2}$ & $2.7 \times 10^{-3}$ & $1.3 \times 10^{-2}$ \\
$p$-value (CP conservation, P odd) & $1.5 \times 10^{-1}$ & $6.9 \times 10^{-2}$ & $6.5 \times 10^{-2}$ \\
$p$-value (P conservation) & $1.3 \times 10^{-7}$ & $4.0 \times 10^{-7}$ & $1.6 \times 10^{-1}$ \\
\hline
\end{tabular}
\caption{The $p$-values from the energy test for different distance scales and test configurations.}
\label{tab:pvalue}
\end{table}

\section{Observation of the suppressed $\Lambda_b^0 \rightarrow D^0 p K^-$ decay with $D^0 \rightarrow K^- \pi^+$ and measurement of its CP asymmetry}

\begin{figure}
    \centering
    \includegraphics[angle=270, scale = 0.4]{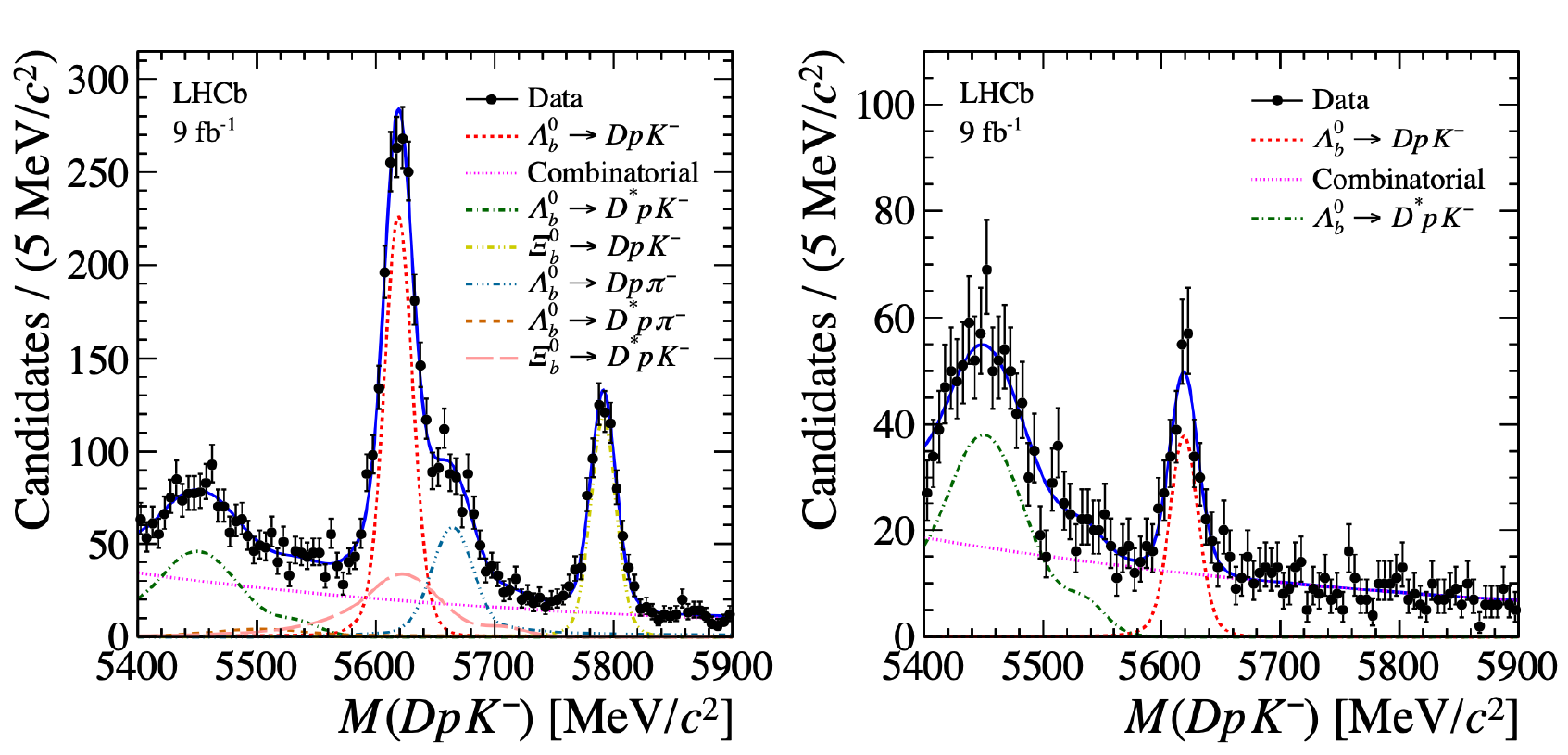}
    \caption{Distributions of the invariant mass for selected (left) $\Lambda_b^0 \rightarrow [K^-\pi^+]_D p K^-$ and (right) $\Lambda_b^0 \rightarrow [K^+\pi^-]_D p K^-$ candidates in the full phase space (black points) corresponding to the favoured and suppressed decays, respectively. The total fit model is indicated by the solid blue line, and individual components are indicated}
    \label{fig:LbD}
\end{figure}

Another study the LHCb collaboration published is the search for the $\Lambda_b^0 \rightarrow D p K^-$, the Cabibbo-favoured and doubly Cabibbo-suppressed mode $D \rightarrow K \pi$, where $D$ represents a superposition of $D^0$ and $\bar{D^0}$ states \cite{Aaij_2021Lb}. In the same study, they performed a measurement of the ratio of the branching fraction of the two decays and the \textit{CP} asymmetry of the suppressed mode. It is possible to perform an estimation of the first quantity via the CKM matrix elements, obtaining
\begin{equation}
    R \approx \left| \frac{V_{cb} V^*_{us}}{V_{ub} V^*_{cs}} \right|^2 = 6.0.
\end{equation}
The $\Lambda_b^0 \rightarrow [K^-\pi^+]_D p K^-$ ($\Lambda_b^0 \rightarrow [K^+\pi^-]_D p K^-$) decay with same (opposite) sign kaons are referred to as the favoured (suppressed) decay throughout this section.
The suppressed decay object of this analysis is of particular interest because its decay amplitude receives contributions almost equally by both $b \rightarrow c$ and $b \rightarrow u$ transition amplitudes, due to the comparable CKM suppression affecting the two $D$ decays. It is expected that the interference of these two amplitudes, which is contingent on the CKM angle $\gamma$, will be large. The data analysed comes from $pp$ collisions at an integrated luminosity of 9 fb$^{-1}$ collected over the period from 2011 to 2018 at center-of-mass energies of $\sqrt{s}$ = 7, 8, and 13 TeV. The result of the selection is showed in Figure \ref{fig:LbD}.
The \textit{CP} measurement is performed both in the full phase space, but also in a restricted one more sensitive to \textit{CP} violation, where the $\Lambda^*$ resonance is predominant, $m^2(pK^-) < 5$ $\text{GeV}^2/c^4$. The \textit{CP} observable is the following
\begin{equation}
    A = \frac{\mathcal{B}(\Lambda_b^0 \to [K^+ \pi^-]_D p K^-) - \mathcal{B}(\bar{\Lambda}_b^0 \to [K^- \pi^+]_D \bar{p} K^+)}{\mathcal{B}(\Lambda_b^0 \to [K^+ \pi^-]_D p K^-) + \mathcal{B}(\bar{\Lambda}_b^0 \to [K^- \pi^+]_D \bar{p} K^+)},
\end{equation}
which has been computed taking into account the per-event efficiency of the selection and the weights obtained using the \textit{sPlot} technique. Also the ratio between the suppressed and the favoured decays is expressed with the same parametrization:
\begin{align*}
A &= \frac{\sum_{i} w^{i}_{\text{SUP},\Lambda^{0}_{b}} / \epsilon^{i} - \sum_{i} w^{i}_{\text{SUP},\bar{\Lambda}^{0}_{b}} / \epsilon^{i}}{\sum_{i} w^{i}_{\text{SUP},\Lambda^{0}_{b}} / \epsilon^{i} + \sum_{i} w^{i}_{\text{SUP},\bar{\Lambda}^{0}_{b}} / \epsilon^{i}} & R &= \frac{\sum_{i} w^{i}_{\text{FAV}} / \epsilon^{i}}{\sum_{i} w^{i}_{\text{SUP}} / \epsilon^{i}}.
\end{align*}
The results are, in the full phase space and in the restricted one:
\begin{align*}
    A &= 0.12 \pm 0.09 \text{ (stat.)}^{+0.02}_{-0.03} \text{ (syst.)} & A &= 0.01 \pm 0.16 \text{ (stat.)}^{+0.03}_{-0.02} \text{ (syst.)} \\
    R &= 7.1 \pm 0.8 \text{ (stat.)}^{+0.4}_{-0.3} \text{ (syst.)} & R &= 8.6 \pm 1.5 \text{ (stat.)}^{+0.4}_{-0.3} \text{ (syst.)}
\end{align*}
The analysis provided to the scientific community a first observation of the suppressed $\Lambda_b^0 \rightarrow [K^+\pi^-]_D p K^-$, and the ratio of the branching fractions of the favoured and suppressed mode, which is, within the uncertainties, compatible with the estimate computed using the relevant CKM matrix elements. The asymmetry values are compatible with zero in the regions studied. 

\section{Search for \textit{CP} violation in $\Xi_b^- \rightarrow p K^-K^-$ decays}
The LHCb collaboration published the first amplitude analysis of a baryon decay allowing for \textit{CP} violation \cite{PhysRevD.104.052010}. The analysis studied the data collected by the LHCb detector from $pp$ collisions, with an integrated luminosity of 5 fb$^{-1}$ collected over the period from 2011 to 2016 at collision energies of $\sqrt{s}$ = 7, 8, and 13 TeV. The amplitude analysis started after the selection performed on the dataset, which allowed to have $(63 \pm 3)\%$ of signal purity in Run I and $(70 \pm 2) \%$ in Run II, considering only candidates in the $m(pK^-K^-)$ signal region of $\pm 40$ MeV around the $\Xi_b^-$ mass. It is assumed that the $\Xi_b^-$ candidates are produced in $pp$ collisions within the LHCb acceptance have negligible polarisation, as observed for the $\Lambda_b^0$. With this assumption it is possible to characterise the phase space of the $\Xi_b^- \rightarrow p K^-_1K^-_2$ decay with only two independent kinematic variables $m^2(pK^-_1)$ and $m^2(pK^-_2)$. Furthermore, to remove the Bose symmetry coming from the exchange of the two $K$, the variables used for the amplitude analysis are instead $m^2_{low}$ and $m^2_{high}$, denoting the lower and the higher range of $m^2(pK^-_1)$ and $m^2(pK^-_2)$ respectively. 
The helicity formalism developed for the analysis relies on the $\Sigma^*$ and $\Lambda^*$ resonances, in the channel $\Xi_b^- \rightarrow (R \rightarrow p K^-) K^-$, such that the differentiel decay density can be expressed as
\begin{equation}
    \frac{d\mathit{\Gamma}^Q}{d\Omega} = \frac{1}{(8\pi m_{\Xi_b})^3} \sum_{M_{\Xi_b}, \lambda_p} \left| \sum_R A^Q_{R,M_{\Xi_b}, \lambda_p}(\Omega) \right|^2 ,
\end{equation}
where $Q = +1$ is for $\Xi_b^-$ and $Q = -1$ is for $\bar{\Xi}_b^+$, and $A^Q_{R,M_{\Xi_b}, \lambda_p}(\Omega)$ denotes the symmetrized decay amplitude for a given intermediate state $R$, $\Xi_b$ spin component along a chosen quantization axis $M_{\Xi_b}$ and proton helicity $\lambda_b$. To formulate the model describing the signal, the resonance $\Lambda(1520)$ has been taken as baseline, leaving the fit free to vary the helicity couplings of all the other resonant and non-resonant components. The chosen model has been obtained when adding further resonances was not resulting in a substantial change in the negative log-likelihood, and is reported in Table \ref{tab:model}. 
\begin{table}[ht]
\centering 
\small
\begin{tabular}{ccccc}
\toprule
State & Mass (MeV/$c^2$) & Width (MeV/$c^2$) & $J^P$ & $A_{CP}(10^{-2})$ \\
\midrule
$\Lambda(1405)$ & $1405.1 \pm 1.3$ & $50.5 \pm 2.0$ & $\frac{1}{2}^-$ & $-27 \pm 34 \,(\text{stat}) \pm 73 \,(\text{syst})$ \\
$\Lambda(1520)$ & $1518$ to $1520$ & $15$ to $17$ & $\frac{3}{2}^-$ & $-1 \pm 24 \,(\text{stat}) \pm 32 \,(\text{syst})$ \\
$\Lambda(1670)$ & $1660$ to $1680$ & $25$ to $50$ & $\frac{1}{2}^-$ & $-5 \pm 9 \,(\text{stat}) \pm 8 \,(\text{syst})$ \\
$\Sigma(1385)$ & $1383.7 \pm 1$ & $36 \pm 5$ & $\frac{3}{2}^+$ & $3 \pm 14 \,(\text{stat}) \pm 10 \,(\text{syst})$ \\
$\Sigma(1775)$ & $1770$ to $1780$ & $105$ to $135$ & $\frac{5}{2}^-$ & $-47 \pm 26 (\text{stat}) \pm 14 (\text{syst})$ \\
$\Sigma(1915)$ & $1900$ to $1935$ & $80$ to $160$ & $\frac{5}{2}^+$ & $11 \pm 26 (\text{stat}) \pm 22 (\text{syst})$ \\
\bottomrule
\end{tabular}
\caption{Components of the model, properties and results for the \textit{CP} asymmetry parameters. }
\label{tab:model}
\end{table}

\begin{figure}[h]
    \centering
    \includegraphics[scale=0.4]{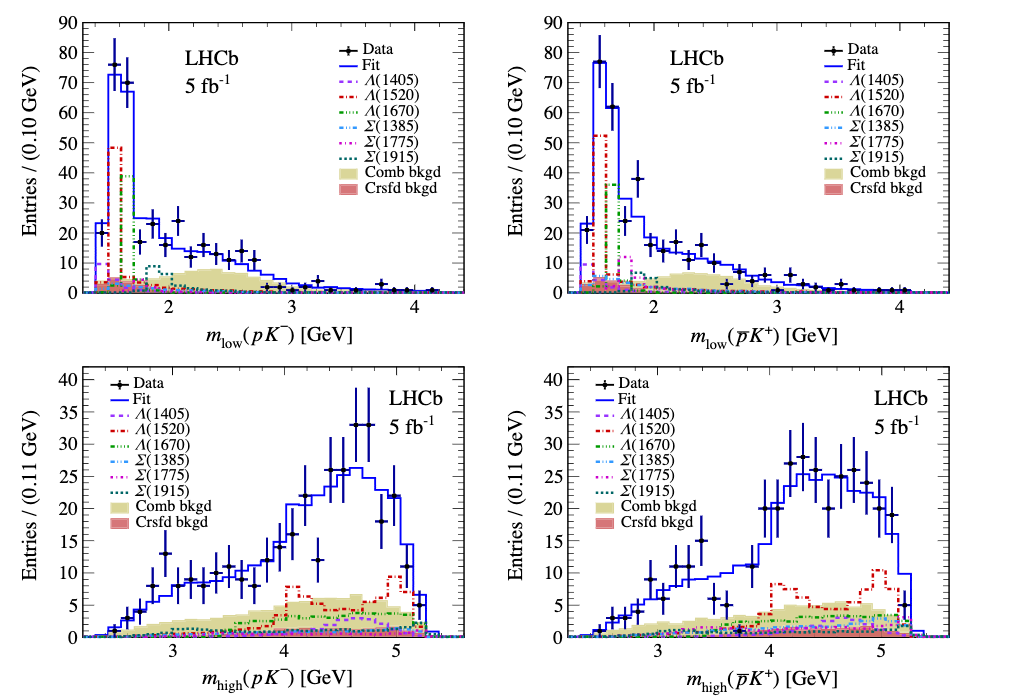}
    \caption{Distributions of (top) $m_{low}$ and (bottom) $m_{high}$ for (left) $\Xi_b^-$ and (right) $\bar{\Xi}_b^+$ candidates, with results of the fits superimposed. The total fit result is shown as the blue solid curve, with contributions from individual signal components and backgrounds.}
    \label{fig:fitXib}
\end{figure}

The fit to data using the selected model is shown in Figure \ref{fig:fitXib}. There is no indication of \textit{CP} violation in the distributions, i.e., no significant difference between $\Xi_b^-$ and $\bar{\Xi}_b^+$ decays.









\bibliography{ref}
\bibliographystyle{unsrt}



\end{document}